\documentclass[fleqn,twoside]{article}
\usepackage{espcrc2}

\usepackage{graphicx}
\usepackage[figuresright]{rotating}

\newcommand{\AmS}{{\protect\the\textfont2
  A\kern-.1667em\lower.5ex\hbox{M}\kern-.125emS}}

\newcommand{\dmd}       {\ensuremath{\mathrm{\Delta {m}_{d}}}}

\newcommand{\ee} {\ensuremath{\mathrm e^+ e^-}}

\newcommand{\bs} {\ensuremath{\mathrm B^0_s}}

\title{Flavor Tagging at Tevatron incl. calibration and control}

\author{T. Moulik (D\O\ Collaboration) \address[MCSD]{Department of Physics and Astronomy,
        University of Kansas, Lawrence, KS 66045}%
        \thanks{MS 352, Fermilab Box 500, Batavia, IL-60510}
.}

\begin{document}

\begin{abstract}
This report summarizes the flavor tagging techniques developed at the CDF
and D\O\ experiments. Flavor tagging involves identification of the $B$ meson flavor at 
production, whether its constituent is a quark or an anti-quark. 
It is crucial for measuring the oscillation frequency of neutral $B$ mesons, 
both in the $B^0$ and $\bs$ system. The two experiments have developed 
their unique approaches to flavor tagging, using neural networks, and likelihood methods
to disentangle tracks from $b$ decays from other tracks. This report
discusses these techniques 
and the measurement of $B^0$ mixing, as a means to calibrate the taggers.
\end{abstract}

\maketitle

\section{INTRODUCTION}
Flavor tagging is important in studies of neutral $B$ meson mixing and CP
violation in the $B$ system. At hadron colliders, because of minimum
bias events and multiple interactions per crossing, 
the background level is higher as compared to the more cleaner
environment at $e^{+} e^{-}$ colliders. Also, at hadron colliders
the $b$-jets may not always be well separated (as in the case of 
$g \rightarrow b \bar{b}$), while at $\ee$ colliders, one gets
well balanced and separated jets. Therefore, in order to identify  $b$-jets or 
decays one needs sophisticated techniques to achieve a good efficiency 
and purity of the tags.  Broadly speaking, flavor tagging methods can be 
categorized into Opposite Side Tagging (OST) and Same Side Tagging (SST) \cite{sst}. 
In the case of OST, one uses the decay products of the opposite $B$ in the event, and since the $B$ mesons are produced in pairs in a $b\bar{b}$ event, the $B$ flavor of the tag side is ideally opposite to the $B$ flavor of the
decay side (The $B$ meson which is fully or partially reconstructed
is referred to as the ``decay'' or ``reconstruction'' side $B$ meson). The tag side $B$ meson however 
could oscillate if it was neutral, independent of the reconstruction side $B$ 
meson and therefore in such a case the tag would incorrect. OST flavor taggers 
are sub-categorized into (i) Lepton taggers, where one exploits the charge correlation
of the lepton and the quark flavor in semileptonic $b$ decays,
(ii) Jet charge taggers, where one use the fact that the kinematically weighted 
charge of the $b$-jet is correlated with the charge of the $b$, and (iii) Opposite Side Kaon Tagger (OSKT), where one exploits the correlation of the
charge of the kaon with the $B$ flavor in the decay chain 
$b \rightarrow c \rightarrow X K^-$.
In the case of SST, one utilizes the charge correlation of the fragmentation
tracks produced on the same side as the ``reconstructed'' $B$. For SST
the wrong sign contribution due to the ``tag'' $B$ meson oscillating is absent,
and it has a high efficiency, but it has a somewhat 
worse purity than the OST tagger, as it is difficult to identify the correct 
fragmentation track from a large number of decay product tracks. 

Flavor taggers are optimized to obtain high efficiencies
($\varepsilon$) and purities, where $\varepsilon$ is defined as the fraction of reconstructed events ($N_{\rm {tot}}$) that are tagged ($N_{\rm {tag}}$):
$\varepsilon = N_{\rm{tag}}/N_{\rm{tot}}$. The term dilution ($\cal{D}$) is
more commonly used, which is related to the purity, $P$, as
${\cal{D}} = 2P - 1$ and is defined as the 
normalized difference of correctly and wrongly tagged events:
${\cal{D}} = \frac{N_{\rm{cor}}-N_{\rm {wr}}}{N_{\rm {cor}}+N_{\rm {wr}}}=\frac{N_{\rm {cor}}-N_{\rm {wr}}}{N_{\rm {tag}}}$
\noindent where, $P=N_{\rm {cor}}/N_{\rm {tag}}$. The 
terms ``correctly'' and ``wrongly'' refer to the determination of the 
decay $B$ meson flavor. The effective tagging power of a 
tagging algorithm is given by $\varepsilon {\cal{D}}^{2}$ and the
goal is to maximize this quantity.

\section{\label{sec:det} CDF AND D\O\ DETECTOR}

Leptons from $B$ decays have low momenta as compared to 
leptons from top or electroweak decays, and require special treatment
to identify them.
The complexity of lepton identification needed depends on the detector.
The CDF detector has a muon coverage up-to $|\eta| \leq 1.5$ and the D\O\ detector 
has a muon coverage up-to $|\eta| \leq 2.0$ in RunII. 
At CDF, the Central muon chamber (CMU) 
is situated just outside the hadron calorimeter.
Hence, the fake rate from punchthroughs, and decay in flights (DIF's) from kaons and pions decaying into muons, for the CMU muons is high. Another layer of muon chambers was therefore installed further up, for the RunII phase, which is acronymed as CMP. 
The rates of punch-through and DIF's are studied in data.
The central muon detectors at 
D\O\ are proportional drift tubes (PDT's) located in the pseudo-rapidity range of
$0.0 < |\eta| < 1.0$ and mini-drift tubes (MDT's) between $1.0 < |\eta| < 2.0$. Both
the central and forward detectors consist of 3 layers, with the first 
layer just outside the hadron calorimeter and the other two layers lying outside the 
toroid. Muons at D\O\ have negligible punchthrough
background after requiring that it passes the toroid. 

The CDF calorimeter is a sampling calorimeter consisting of layers
of scintillators and lead. There is a position detector at shower maximum, 
at about 6 radiation lengths ($X_0$), which is composed of orthogonal strips and wires, 
and provides measurement in the $x$ and $z$ direction.
Besides this, a central pre-radiator at about 1 $X_0$ is 
useful as an additional layer for identifying electrons, especially low momentum electrons 
which can start showering much before the EM calorimeter. This was a gas wire chamber 
in the RunIIa phase and is now replaced by scintillators in the RunIIb phase. 
The D\O\ calorimeter is a primarily liquid-argon/uranium sampling calorimeter, 
covering up-to $|\eta| < 4$.
The central preshower situated at $\approx$ 1 $X_0$, is made of three concentric cylindrical
layers of triangular cross-section scintillator strips, with wavelength shifting 
fiber (WLS) readout.

More details on the CDF 
and D\O\ detector can be found in ref. \cite{cdfdet} and \cite{d0det} respectively.

\section{\label{sec:ost} OPPOSITE SIDE TAGGING}
\subsection{\label{sec:lepid}Soft lepton identification}
\begin{figure}[ht]
\begin{center}
    \includegraphics[width=6.0cm]{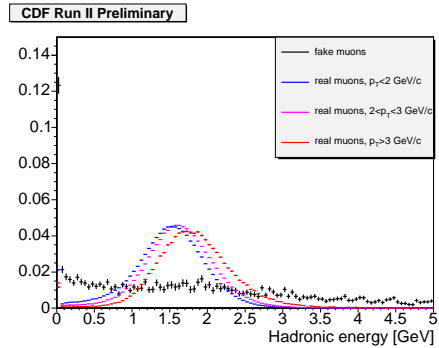}
    \caption[]{
    \label{fig:muonid1} Energy deposit in the hadron calorimeter for
fakes and signal muons at CDF. The black dots come from fakes peaking
at 0, while real muons deposit around 1.5-2.0 GeV depending on their $p_T$,
consistent with a minimum ionizing particle.}
\end{center}
\end{figure}

For identifying muons, CDF developed a muon likelihood function, 
to disentangle background coming from punchthroughs and DIF's, 
which is especially problematic in the CMU. The muon ID variable probability density functions 
(pdf's) were developed from data, using signal muons from 
$J/\Psi \rightarrow \mu^+ \mu^-$ and fakes from $K^0_S \rightarrow \pi^+ \pi^-$ and other modes. The 
energy deposit in the hadron calorimeter for signal and fakes
can be seen in Fig. \ref{fig:muonid1}. 

At D\O, simple cuts are used, as there is negligible punchthrough
after requiring that the muon must pass the toroid.

\begin{figure}[ht]
\begin{center}
\includegraphics[height=5.0cm]{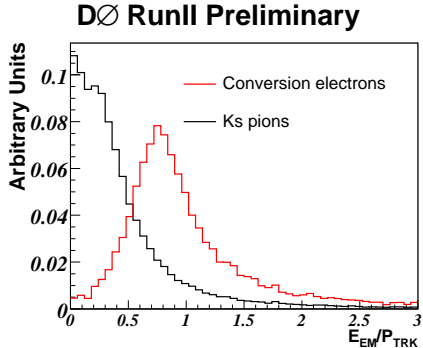}
\end{center}
\caption{\label{fig:eop}
The fraction of energy deposited in the electro-magnetic
calorimeter to the momentum of the track.
}
\end{figure}

The standard electron reconstruction is based on calorimeter
clusters, relying on the characteristic transverse and
longitudinal shapes of electromagnetic showers and is usually designed 
for high $p_T$ electrons. The leptons in $B$ decays have low momenta and could 
be non-isolated and within the b-jet. For low $p_T$ electrons, in order to achieve
a higher purity and to remove contamination from nearby tracks, one starts
with the track and extrapolates it to the calorimeter and it suffices
to use a fewer number of towers to form a narrower cluster.
D\O\ uses simple cuts on the electron ID quantities while CDF uses
a likelihood variable. To optimize the cuts or to develop the likelihood,
electrons from photon conversions $\gamma \rightarrow \ee$ as signal and 
$K^0_S \rightarrow \pi^{+} \pi^{-}$ are used for the fakes.
Electrons from photon conversions have a very similar $p_T$ spectrum as electrons
from $b$-decays. The ratio of energy deposited in the electromagnetic
calorimeter to the momentum of the track for the D\O\ experiment can be see 
in Fig. \ref{fig:eop}

\subsection{OST development at CDF}
\begin{figure}[ht]
\begin{center}
\includegraphics[width=5.7cm]{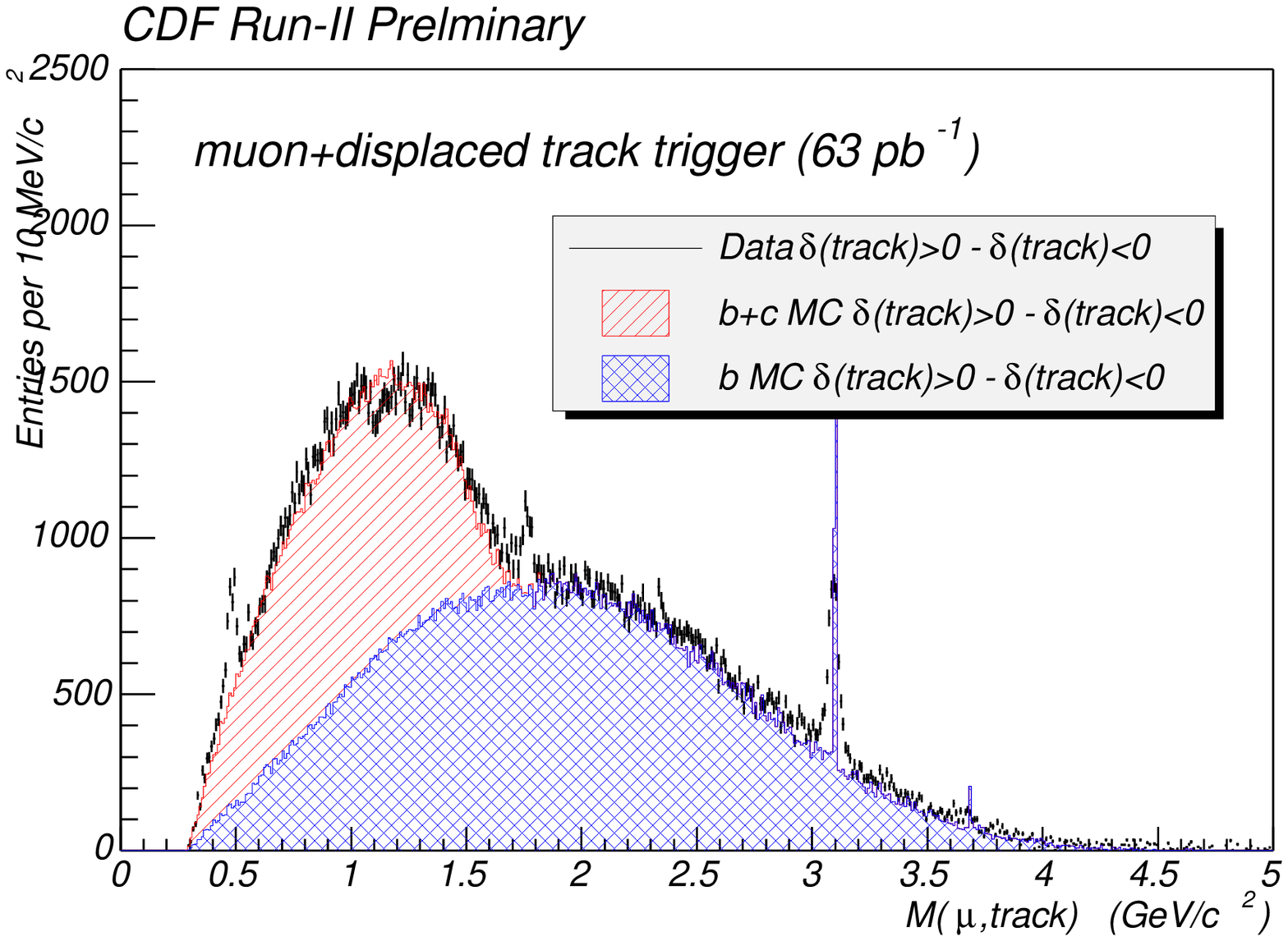} 
\caption{\label{fig:mlsvt}
   The invariant mass distribution for $l+SVT$ sample with the negative impact
parameter background component subtracted. The $b$ and $c$ components from 
MC are also superimposed}
\end{center}
\end{figure}

CDF uses a lepton and displaced track trigger, called Silicon Vertex Trigger (SVT)
at level2 to collect a $b$-enriched inclusive, high statistics sample to
calibrate the lepton taggers. To enhance the $b$ component further, 
the invariant mass of the lepton and displaced 
track system $M_{l+SVT}$ is required to be consistent with a $B$ candidate, 
using the cut, $2 < M_{l+SVT} < 4$ GeV. Furthermore, true $B$ decays should have a 
positively displaced signed impact parameter $\delta$, where 
$\delta = |d_0| sign(d_0 \cdot p_{l+SVT})$, $d_0$ being the impact parameter of the track.
Therefore, a background subtraction is done, and the 
sample with negative impact parameter is subtracted from the
sample with positive signed impact parameter (See Fig. \ref{fig:mlsvt}).
However, since this sample is inclusive of all $B$ decays, the
trigger side can mix and undergo sequential decays and there are also fake
leptons, which means that the sign of the lepton does not give the correct
$B$ meson flavor at the decay side. Thus, the ``true'' tagger dilution is given by 
the ``raw'' tagger dilution/''trigger'' side dilution. 

\subsubsection{Lepton tagging}\label{cdflep}
The sign of the $B$ flavor on the ``decay'' side is given by
the trigger lepton in the $l+SVT$ sample. Tag leptons (muon or electron) 
with $p_T > 2 GeV$ are searched for, on the opposite side using the soft lepton 
identification algorithms as described in section \ref{sec:lepid}. 
Only electrons in the range $|\eta| \leq 1$ are used. The dilution is then
parametrized as a function of the transverse momentum of the tag lepton 
w.r.t the tag lepton jet ($p_T^{\rm rel}$) 
and the electron or muon likelihood cut, to provide an event-by-event dilution \cite{cdfresults}.  

\subsubsection{Jet charge tagging}\label{jqt}
\begin{figure}[ht]
\begin{center}
    \includegraphics[width=4.5cm]{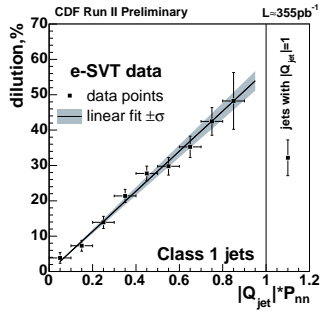}
    \caption[]{
    \label{fig:jettag}  
The dilution as a function of $|Q_{jet}| * P_{nn}$ for class 1 jets in the electron+displaced track trigger sample. 
    }
\end{center}
\end{figure}

The jet charge tag starts with forming track based jets. Each of the tracks
in the jet are assigned a ``b''-ness probability using a neural network,
called a TrackNet probability. Some of the track properties used as 
inputs to the neural network are the impact parameter significance, 
the signed impact parameter, the transverse momentum, and others.
To select the single best $b$-jet in the presence of multiple jets, the jets 
are fed to a neural network, and the jet with the highest NN probability ($P_{nn}$) 
is chosen.
The jet charge is calculated as $Q_{jet} = \frac{\sum q_i p_T^i(1+t_i)}{\sum p_T^i(1+t_i)}$, 
where $t_i$ is the NN probability for the track. The jets are then divided
into three categories based on the presence of a secondary vertex or the number of 
tracks with high NN prob., and for each class the dilution is derived as a function
of $|Q_{jet}| * P_{nn}$. The
NN probability for all the jets and the dilution calibration for class 1
jets can be seen in Fig. \ref{fig:jettag}. The results are summarized
in table \ref{tab:flavtag}.

\subsubsection{Opposite side kaon tagging}
CDF utilizes dE/dx in the tracker and time-of-flight (TOF) information 
to identify kaons. The following results are found for OSKT \cite{cdfresults}:
$\varepsilon  =  18.11 \pm 0.07 \%$ and $\varepsilon {\cal{D}}^2  =  0.229 \pm 0.016 \pm 0.001(\%)$

\subsection{OST development at D0}\label{tagd0}
\begin{figure}[ht]
\begin{center}
    \includegraphics[width=4.0cm]{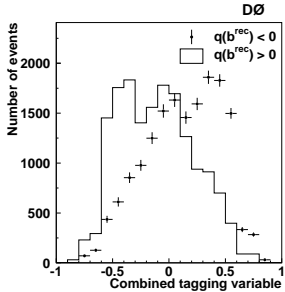}
    \caption[]{
    \label{ftag:fig3}  
	Normalized distributions of the combined tagging variable $d$.
        $q(b^{\rm rec})$ is the
        charge of the $b$ quark from the reconstruction side.
    }
\end{center}
\end{figure}

At D\O\ one collects samples using an inclusive muon trigger.
Then, $B^{+} \rightarrow D^0 \mu^{+} \nu X$ decays are reconstructed
and used for the development of the tagger. 
(Charge conjugated states are implied throughout the note). 
In this sample, the muon charge gives the 
flavor of the reconstruction side $B$ since charged $B$'s do not oscillate. 
This sample has a small contribution from $B^0$ decays but by
requiring a small decay length of the $B$ candidate, the sample is composed of $98 \%$
of the decays from $B^{+}$ mesons. On the opposite side then, one can construct
p.d.f's for tag variables that distinguish between $b$ and $\bar{b}$, the
flavor information being given by the sign of the muon in this sample's case.
For n discriminating variables, therefore the combined tagging variable 
$r$ is defined as:  $r = \prod_{i=1}^{n}r_i$, where $r_i$ is given 
by the ratio of the p.d.f's for a $b$ and a $\bar{b}$ quark.
A more convenient tagging variable is defined as: $ d = \frac{1-r}{1+r}$, which
ranges between $-1$ and $1$, and an event with $d>0$ is tagged as a $b$ quark and with $d<0$ as a 
$\bar{b}$ quark. A  higher $|d|$ value corresponds to higher $b$-ness of
the tag. Specifically, for each event with an identified muon on the opposite side, 
the discriminating variables, muon jet charge $Q_J^\mu$, and the secondary vertex charge $Q_{SV}$ are used to construct 
a {\it muon tagger}. For each event without a muon but with an identified electron, 
the electron jet charge $Q_J^e$ and the secondary vertex charge $Q_{SV}$ are used to construct 
an {\it electron tagger}. Finally, for events without a muon or an electron but with a reconstructed secondary vertex, 
the secondary vertex charge $Q_{SV}$, and the event jet charge $Q_{EV}$ are used to construct 
a {\it secondary vertex tagger}. The resulting distribution of the tagging variable $d$ for the combination 
of all three taggers, called the {combined tagger}, is shown in Fig. \ref{ftag:fig3}.
More details can be found in Ref. \cite{bdmixd0}.

\section {SAME SIDE TAGGING AT CDF}
To develop the same side tagging, especially in the case of $\bs$ mesons, one
has to rely on Monte Carlo (MC) since the tagging is species dependent and in the
case of $B^0$ the fragmentation track is a pion while in the case of $\bs$
its a kaon. Hence, one needs to achieve a good data-MC agreement in the high statistics $B^0$ and $B^{+}$ modes, and
then study the systematic uncertainties of the prediction of the tagger
performance from MC samples for the $\bs$ data. 
Of importance to this analysis
is the use of particle identification systems like the TOF detector, and the use of energy loss (dE/dx) in the drift chamber to 
distinguish between kaon and pion tracks, in particular. 
An effective tagging power of $4.0^{+0.9}_{-1.2} \%$ is found \cite{cdfresults}. 

\begin{figure}[ht]
\begin{center}
    \includegraphics[width=5.0cm]{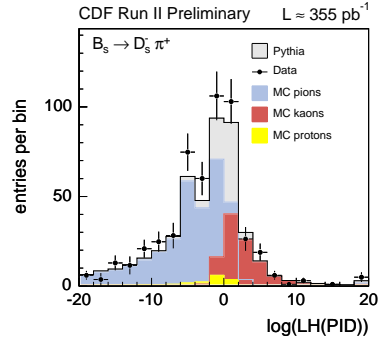}
    \caption[]{\label{fig:sstdatamc}
	Data-MC comparison in $\bs$ decays
for kaon PID variable.
where $log(LH(PID)) = log \frac{P_{TOF} (K) P_{dEdx} (K)}{ {f_{p}} {P_{TOF}} (p) {P_{dEdx}} (p) + {f_{\pi}}{P_{TOF}}(\pi) {P_{dEdx}}(\pi) }$ and $f_p = 0.1$ and $f_\pi = 0.9$ are a priori probabilities of background composition and $P_{TOF}$ and $P_{dE/dx}$ are determined from pure samples. 
}
\end{center}
\end{figure}

\section {$B^0$ MIXING AND TAGGER CALIBRATION}
$B^{+} \rightarrow \bar{D}^0 \mu^{+} X$ and $B^0 \rightarrow D^{*-} \mu^{+} X$ samples are used for measuring $B^0$ mixing 
and for tagger calibration. The sample composition of $B$ meson species and $c\bar{c}$ events contributing to the final states 
are obtained from realistic simulated events. 

\subsection{Results from D\O}
\begin{figure}[ht]
    \includegraphics[width=7.5cm]{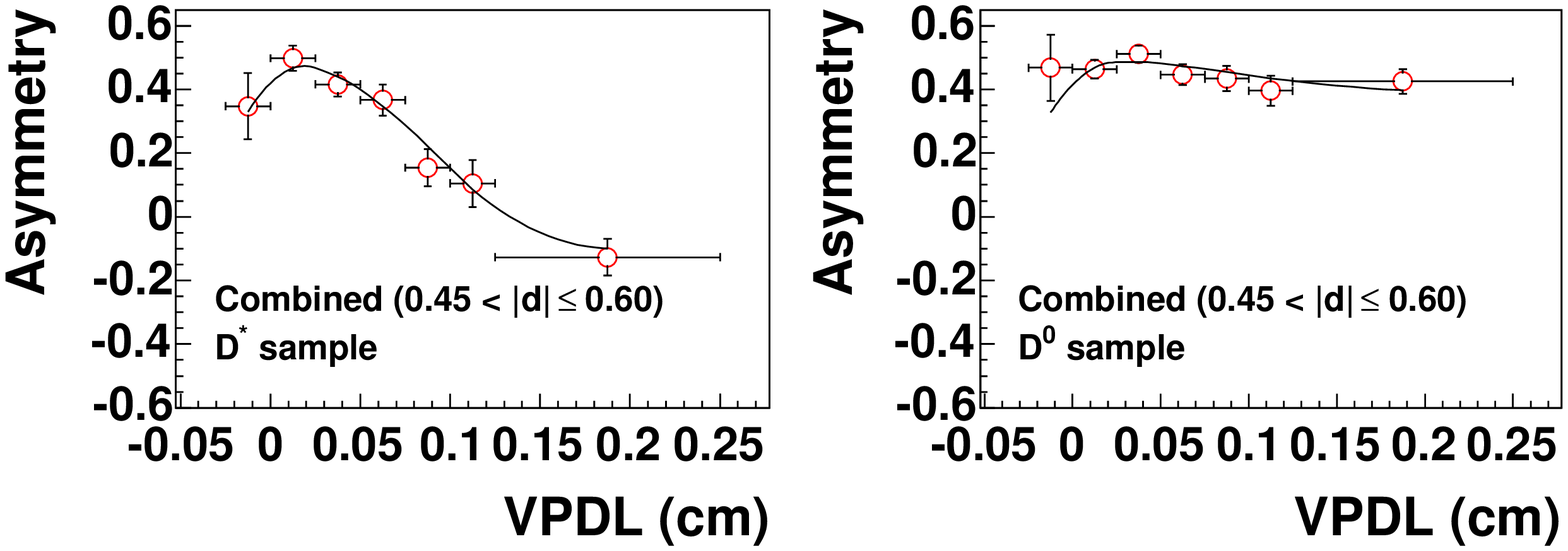}
    \caption[]{
    \label{fig:tagd01} 
Asymmetries obtained in the $D^0$ and $D^{*}$ data samples with the combined tagger as a function of visible proper decay length (VPDL) in 
one of the $|d|$ bins at the D\O\ experiment. The circles are data and the fit is superimposed.
    }
\end{figure}

The asymmetries given by the difference of opposite-sign and same-sign
tagged events are obtained by fitting the $D^0$ mass distribution in the
$D^0$ and $D^{*}$ samples as a function of the visible proper 
decay length ($x^{M}$) of the D+$\mu$ candidate. 
$x^{M}$ is related to the real $B$ decay length as $K x^M = c \tau_B$, where K-factor is evaluated from MC as the ratio of the transverse momenta of the $D+\mu$ candidate and the generated B meson.
The asymmetries can then be fitted in a binned $\chi^2$ fit to extract 
$\dmd$ and the tagger dilution ${\cal{D}}$. The samples 
are divided into 5 sub-samples of low to high purity in the tag variable $|d|$ and fitted in a combined fit to 
extract the ${\cal{D}}$'s as a function of $|d|$ thus providing an event-by-event dilution. 
This calibration curve is then used for $\bs$ mixing studies. The asymmetry fit in one of the 5 $|d|$ bins can be seen in Fig. \ref{fig:tagd01}. More details can be found in Ref. \cite{bdmixd0}. Summing the individual tagging powers of all $|d|$ bins after the fit, one obtains an effective tagging power of $\varepsilon {\cal D}^2=[{2.48 \pm 0.21~{\rm (stat.)}}^{+0.08}_{-0.06}~{\rm(syst)}]~ \%$.
The fraction $f_{c \bar c}$ which is constrained to be the same for all subsamples is found to be $f_{c \bar c}  =  (2.2 \pm 0.9) \%$, and
the $B^0$ mixing parameter $\dmd$ is found to be $\dmd  =  0.506 \pm 0.020(stat.) \pm 0.016 (syst.) ~\mbox{ps}^{-1}$ in good agreement with world average value of $0.507 \pm 0.005 ~{\rm ps}^{-1}$ \cite{pdg2006}.

\subsection{Results from CDF}
CDF obtains the dilution calibration from the inclusive lepton+SVT samples (See Sec.\ref{cdflep} and \ref{jqt}) and introduces a scale factor to describe the differences in dilution for the individual decay
modes considered. Using a sample corresponding
to an integrated luminosity of $355 \,{\rm pb}^{-1}$, an unbinned likelihood simultaneous fit to both 
mass and lifetime is performed and a $\dmd$ of
$0.536 \pm 0.028 \pm 0.006~\mbox{ps}^{-1}$ and an $\varepsilon {\cal{D}}^2  =  (1.55 \pm 0.16 \pm 0.05) \%$ is found in the case of hadronic decays, and using a similar method and an integrated luminosity of
$1 \, {\rm fb}^{-1}$, a $\dmd  =  0.509 \pm 0.010 \pm 0.016~\mbox{ps}^{-1}$ and
$\varepsilon {\cal{D}}^2 =  (1.54 \pm 0.05 ) \%$ is obtained in the case of semileptonic
decays \cite{cdfresults}.


\section{FLAVOR TAG SUMMARY}
Overall tagging performances for the two experiments is summarized in Table \ref{tab:flavtag}.
For the individual tagger numbers quoted for D\O\ a cut of $|d|>0.3$ was used. The D\O\ combined
tagger uses events with $|d| < 0.3$, hence overall effective tagging power is higher. 
CDF also used a neural network method to combine the individual taggers and finds
an increase in OST effective tagging power from $1.5 \%$ to  $1.8 \%$ \cite{cdfresults}.
After demonstrating a consistent measurement of $B^0$ mixing, the
flavor tagging calibration can then be used for $\bs$ mixing studies and
was used to obtain the $\bs$ mixing results at the two experiments
\cite{bscdf}, \cite{bsd0}.

\begin{table}[htb]
\begin{center}
\caption[]{\label{tab:flavtag}Flavor Tag Summary at the Tevatron. Here,
the soft muon and soft electron taggers are denoted SMT and SET respectively.
The three categories of Jet Charge Taggers (JQT) (See Sec. \ref{jqt}), 
jets with an associated secondary vertex, highly displaced tracks content and the
rest of the jets, are referred to as JVX, JJP and JPT.
The JVX tagger at CDF is similar to the secondary vertex tagger developed at 
D\O\ experiment(Sec. ~\ref{tagd0}).}
\newcommand{\m}{\hphantom{$-$}}
\newcommand{\cc}[1]{\multicolumn{1}{c}{#1}}
\renewcommand{\tabcolsep}{0.5pc} 
\renewcommand{\arraystretch}{1.2} 
\begin{tabular}{@{}|l|lll|lll|}
\hline
& \multicolumn{3}{|c|}{CDF} 
& \multicolumn{3}{|c|}{D\O} \\\hline
Tagger                    
& \m$\varepsilon$ 
& \m$\cal{D}$
& \m$\varepsilon {\cal{D}}^2 $
& \m$\varepsilon $
& \m$\cal{D}$
& \m$\varepsilon {\cal{D}}^2 $ \\
       & $(\%)$  &        & $(\%)$  & $(\%)$ &      &  $(\%)$ \\
\hline
SMT     & \m4.8  & \m0.36 & \m0.54  & \m6.6 & \m0.47 & \m1.48 \\
SET     & \m3.1  & \m0.30 & \m0.29  & \m1.8 & \m0.34 & \m0.21 \\
JVX     & \m7.7  & \m0.20 & \m0.23  & \m2.8 & \m0.42 & \m0.50 \\
JJP     & \m11.4 & \m0.11 & \m0.35  & ...   &  ...   & ...    \\
JPT     & \m57.9 & \m0.05 & \m0.09  & ...   &  ...   & ...    \\\hline
OST     & \m94.7 &        & \m1.50  & \m19.0 &       & \m2.5 \\\hline
SST     & \m54.0 & \m28.3 & \m4.00  & ... & ... & ... \\
\hline
\end{tabular}
\end{center}
\end{table}


\end{document}